**Univariate Likelihood Projections and Characterizations of the Multivariate Normal Distribution**

Albert Vexler

*Department of Biostatistics, The State University of New York at Buffalo, Buffalo, NY 14214, U.S.A*

*avexler@buffalo.edu*

The problem of characterizing a multivariate distribution of a random vector using examination of univariate combinations of vector components is an essential issue of multivariate analysis. The likelihood principle plays a prominent role in developing powerful statistical inference tools. In this context, we raise the question: can the univariate likelihood function based on a random vector be used to provide the uniqueness in reconstructing the vector distribution? In multivariate normal (MN) frameworks, this question links to a reverse of Cochran's theorem that concerns the distribution of quadratic forms in normal variables. We characterize the MN distribution through the univariate likelihood type projections. The proposed principle is employed to illustrate simple techniques for assessing multivariate normality via well-known tests that use univariate observations. The presented testing strategy can exhibit high and stable power characteristics in comparison to the well-known procedures in various scenarios when observed vectors are non-MN distributed, whereas their components are normally distributed random variables. In such cases, the classical multivariate normality tests may break down completely.

## 1. INTRODUCTION

In various theoretical and applied studies, multivariate analysis treats multivariate normally distributed data (e.g., Kotz et al., 2000). There is an extensive amount of fundamental results related to characterizations of the multivariate normal distribution. In this context, characterizations of multivariate normality (MN) through univariate projections play fundamental roles, providing relatively simple procedures to assess the assumption of MN regarding a random vector distribution (e.g., Shao and Zhou, 2010; Cuesta-Albertos et al., 2012; Looney, 1995). Perhaps, mostly addressed univariate characterization of MN employs that the random variables $X_1,\ldots,X_p$ are jointly normal if and only if every linear combination of them is a univariate normal. This property underlies many strategies of testing for MN that have structures of powerful techniques developed in the univariate cases (e.g., Looney, 1995; Zhu et al., 1995).

An important critical result is that the MN of all subsets $(r < p)$ of the normal variables $X_1,\ldots,X_p$ together with the normality of an infinity number of linear combinations of them do not insure the joint normality of these variables (e.g., Hamedani, 1984). This raises a vital concern regarding the common statistical procedures for assessing MN of a random vector by examining a limited number of linear combinations of its components (e.g., Shao and Zhou, 2010). In practice, technical reasons restrict the number of the linear combinations to be considered.

In this paper, we introduce an alternative univariate projection of MN that is inspired by the following statements. The likelihood principle plays a prominent role in developing powerful statistical inference tools (e.g., Vexler and Hutson, 2018). Oftentimes, likelihood functions assist to derive sufficient information regarding observed data. Then, one might ask: can a distribution of the likelihood function based on the vector $X = (X_1,\ldots,X_p)^T$ be involved in complete reconstruction of $X$'s distribution? The likelihood function based on $X$ is a univariate random variable.

In the case where $X$ is MN distributed, the corresponding log likelihood function can be directly associated with so called quadratic forms (see Section 2 for details). According to Ruben (1978), "from a substantive or statistical point of view the characterization of normality via quadratic forms must rank as of greater interest when one bears in mind that the core of statistical science, namely the entire vast area of regression analysis, including analysis of variance, is based on quadratic forms of the components of the observation vector." Ruben (1978) provided basic characterizations of normality, showing that, when $X_1,\ldots,X_p$ are symmetric, independently and identically distributed random variables with zero means and finite variances, the corresponding quadratic form has a chi-squared distribution if and only if $X_1$ is normal. This approach can characterize $X_1,\ldots,X_p$ as normally distributed random variables, but does not sufficiently imply that $X \sim MN$ (Hamedani, 1984). Indeed, it is of theoretical and applied interest to release the conditions regarding independence of $X$'s components and their symmetry.

In Section 2, we establish a new characterization of MN for a random vector by examining the relevant quadratic form. The obtained results can underlie a reverse of Cochran's theorem (e.g., Styan, 1970) that concerns the distribution of quadratic forms in normal variables. It turns out that, in general cases, we can provide one-to-one mapping between the likelihood's and $X$'s distributions, using properties of infinity divisible (ID) distribution functions. For an extensive review and examples related to univariate and multivariate ID distributions, we refer the reader to Bose, Dasgupta and Rubin (2002). We point out that the problem of univariate likelihood projections can be linked to the issue of reconstructing summands distributions by a distribution of their sum. In this context, the conclusions of Prokhorov and Ushakov (2002) (see Theorem 1 and its Corollary in the cited paper) show that, even in the simple case of independent $X_1,\ldots,X_p$, the ID assumption applied in Section 2 cannot be significantly improved.

In Section 3, we exemplify an application of the proposed method, constructing simple tests for MN. Although many techniques for assessing MN have been proposed, there is still a paucity of genuine statistical tests for MN (e.g., Kotz et al., 2000). Taking into account the arguments presented by Looney (1995), we introduce techniques for assessing MN based on well-known tests that use univariate observations. We experimentally show that the proposed likelihood projection based testing strategy can exhibit high and stable power characteristics in comparison to the relevant well-known classical procedures in various scenarios when $X$ is not MN-distributed, whereas $X_1,\ldots,X_p$ are dependent or independent normally distributed random variables (Stoyanov, 2014). In such cases, the Shapiro-Wilk, Henze-Zirklers and the Mardia multivariate normality tests may break down completely. We conclude with remarks in Section 4.

## 2. LIKELIHOOD PROJECTIONS

We first introduce the basic notation regarding the statement of the problem. Then the main results are provided in Theorems 1 and 2 that establish univariate likelihood based characterizations of MN. The proofs of Theorems 1 and 2 are included for completeness and contain comments that assist to describe the obtained results. Important notes related to conditions used in the proposed technique are presented in Remarks.

Let $X = (X_1,\ldots,X_p)^T$ be the $p$-dimensional random vector with mean vector $\mu = (\mu_1,\ldots,\mu_p)^T$ and covariance matrix $\Sigma$. The covariance matrix $\Sigma$ is positive-definite. Then we can use an orthogonal matrix $Q$ to present the diagonalizable form of $\Sigma$, $Q^T \Sigma Q = \Lambda$, where

$$\Lambda = \begin{bmatrix} \lambda_1 & 0 & \cdots & 0 \\ 0 & \lambda_2 & \cdots & 0 \\ \vdots & \vdots & \vdots & \vdots \\ 0 & 0 & \cdots & \lambda_p \end{bmatrix}, \quad \lambda_i > 0, i = 1,\ldots,p,$$

(e.g., Baldessari, 1967). Define the following matrices

$$\Delta = \begin{bmatrix} 1/\lambda_1^{1/2} & 0 & \cdots & 0 \\ 0 & 1/\lambda_2^{1/2} & \cdots & 0 \\ \vdots & \vdots & \vdots & \vdots \\ 0 & 0 & \cdots & 1/\lambda_p^{1/2} \end{bmatrix}, \quad I = \begin{bmatrix} 1 & 0 & \cdots & 0 \\ 0 & 1 & \cdots & 0 \\ \vdots & \vdots & \vdots & \vdots \\ 0 & 0 & \cdots & 1 \end{bmatrix}, \quad H = Q^T \Delta Q, \text{ and } z = H(X - \mu).$$

Obviously $H$ is symmetric (e.g., $H = H^T$) and $H^T H = Q^T \Delta Q Q^T \Delta Q = Q^T \Lambda^{-1} Q = Q^T \Sigma^{-1} Q$, since the equation $Q^T \Sigma Q = \Lambda$ provides $\Sigma = Q \Lambda Q^T$ after applying the inverse of both sides and using that the inverse of an orthogonal matrix is equal to its transpose. Also we have

$$H \Sigma H^T = Q \Delta Q^T \Sigma Q \Delta Q^T = Q \Delta Q^T Q \Lambda Q^T Q \Delta Q^T = I.$$

Assuming that $X$ is observed and follows a multivariate normal distribution, say $X \sim N_p(\mu, \Sigma)$, we can write the conventional likelihood function $L = (2\pi)^{-p/2} |A|^{1/2} \exp\{-0.5W\}$, where $A = \Sigma^{-1}$ is a real symmetric positively defined matrix and the quadratic form $W = (X - \mu)^T A (X - \mu)$. It is clear that the distribution of $W$ determines the distribution of $L$ and vice versa. Note that

$$W = (X - \mu)^T \Sigma^{-1} (X - \mu) = (X - \mu)^T H^T H (X - \mu) = [H(X - \mu)]^T H(X - \mu) = z^T z.$$

**Theorem 1 (Likelihood Projection and Characterization).** The following two statements are equivalent:

(a) $X$ is an infinitely divisible random (ID) vector, the random vector $z$ consists of $p$ independent components and the random variable $z^T z = W$ has the chi-square distribution with $p$ degrees of freedom, say $z^T z \sim \chi_p^2$.

(b) $X \sim N_p(\mu, \Sigma)$

**Proof.** Under Statement (a), we have $z = H(X - \mu) = (z_1, ..., z_p)^T$, where $z_i$ is a linear combination of $X_1, ..., X_p$, $i = 1, ..., p$. Therefore, for all $i = 1, ..., p$, $z_i$ is an ID random variable (e.g., Horn and Steutel, 1978: Theorem 3.2; Rao, 2012: p. 66). (Note that, in this case, the assumption: "$X$ is an ID random vector" is employed, whereas, in general, a linear combination of ID random variables can be not an ID random variable. Here, for example, in a particular case, we can regard a structure of the definition of normally distributed random vectors, comparing to that of normally distributed random variables, and refer to, e.g., Hamedani. 1984.) Then we apply the following result, focusing on $z^T z = z_1^2 + ... + z_p^2$, where $z_1, ..., z_p$ are independent.

**Proposition 1** (Kruglov, 2013). If $Y_1, ..., Y_p$ are independent ID random variables such that $Y_1^2 + ... + Y_p^2$ has the chi-square distribution with $p$ degrees of freedom then random variables $Y_1, ..., Y_p$ have the same standard normal distribution.

Thus, for all $i = 1, ..., p$, $z_i \sim N_1(0,1)$. Since $(X - \mu) = H^{-1} z$, for all $i = 1, ..., p$, $X_i - \mu_i$ is an linear combination of independent identically $N_1(0,1)$-distributed $z_1, ..., z_p$. Then, the simple use of a characteristic function of $X_i - \mu_i$ shows that $X_i - \mu_i \sim N_1$, $i = 1, ..., p$. (Note that, in this case, we use that $z_1, ..., z_p$ are independent and identically distributed, whereas, in general, a linear combination of normally distributed random variables can be non-normally distributed.) Now, Propositions 1 and 2 of Wesolowski (1993) assist to conclude that the ID random vector $X \sim N_p(\mu, \Sigma)$ that is Statement (b).

Under Statement (b), it is clear that $X$ is an ID random vector and we have the quadratic form $z^T z = W = (X - \mu)^T A (X - \mu) \sim \chi_p^2$ by virtue of Cochran's theorem (e.g., Styan, 1970). In this case, $E(z) = H[E(X - \mu)] = 0$ and $\text{var}(z) = H \text{var}(X - \mu) H^T = H \Sigma H^T = I$, and then $z \sim N_p(0, I)$, since $X \sim N_p(\mu, \Sigma)$. These provide Statement (a) and then we complete the proof.

**Example:** Consider the bivariate scenario $p = 2$, where $E(X_i) = \mu_i$, $\text{var}(X_i) = 1$ and $\text{cor}(X_1, X_2) = \rho$, $i = 1, 2$. In this case,

$$z_1 = h_1(X_1 - \mu_1) + h_2(X_2 - \mu_2) \text{ and } z_2 = h_2(X_1 - \mu_1) + h_1(X_2 - \mu_2),$$

where $h_1 = 0.5\left[(1-\rho)^{1/2} + (1+\rho)^{1/2}\right] / (1-\rho^2)^{1/2}$ and $h_2 = 0.5\left[(1-\rho)^{1/2} - (1+\rho)^{1/2}\right] / (1-\rho^2)^{1/2}$.

Theorem 1 states that $X \sim N_2$ if and only if (iff) $X$ is an ID random vector,

$$W = z^T z = \frac{(X_1 - \mu_1)^2}{(1-\rho^2)} + \frac{(X_2 - \mu_2)^2}{(1-\rho^2)} - 2\rho \frac{(X_1 - \mu_1)(X_2 - \mu_2)}{(1-\rho^2)} \sim \chi^2_{p=2}$$

and the random variables $z_1$, $z_2$ are independent. Note also that, if $\tilde{X} = (\tilde{X}_1, \tilde{X}_2)^T \sim N_2$ with $E(\tilde{X}_i) = \tilde{\mu}_i$, $\text{var}(\tilde{X}_i) = \sigma_i^2$ and $cor(\tilde{X}_1, \tilde{X}_2) = \rho$, $i = 1, 2$, the transformed vector

$X = \left[(\tilde{X}_1 - \tilde{\mu}_1)/\sigma_1, \ (\tilde{X}_2 - \tilde{\mu}_2)/\sigma_2\right]^T$ can be evaluated in the manner shown above, when $\mu_i = 0$.

In this case, the matrix $H = \begin{bmatrix} h_1 & h_2 \\ h_2 & h_1 \end{bmatrix}$ is symmetric.

**Remark 2.1.** It seems that the ID requirement used in Theorem 1 can be substituted by a symmetric type restriction on $z$'s distributions (see the Introduction of Kruglov, 2013 as well as Ruben, 1978). This approach leads to characterize $X_1, \ldots, X_p$ as $N_1$-distributed random variables, but cannot sufficiently assist to conclude that $X \sim N_p(\mu, \Sigma)$ (Hamedani, 1984). This is one of reasons to require that $X$ is an ID vector. In this case the ID restriction on $z$'s distributions is more profound than the symmetric distributions' considerations (Kruglov, 2013: p. 873).

**Remark 2.2.** A set of results regarding situations when ID vectors are normally distributed can be found in, e.g., Wesolowski (1993) and Bose et al. (2002: p. 783). Bose et al. (2002) provided an extensive review and examples related to ID distributions.

The following proposition can get involved into the Theorem 1 structure instead of Proposition 1.

**Proposition 2** (Golikova and Kruglov, 2015). Let $Y_1, \ldots, Y_p$, $p \geq 2$ be independent ID random variables. The random variable $\sum_{i=1}^{2}\left(Y_i - \sum_{j=1}^{2} Y_j / 2\right)^2$ has the chi-square distribution with 1 degree of freedom iff $Y_1$ and $Y_2$ are Gaussian random variables with $EY_1 = EY_2$ and $E(Y_1 - EY_1)^2 + E(Y_2 - EY_2)^2 = 2$. In general for $p \geq 3$, if $EY_1 = \cdots = EY_p$ and the random variable $\sum_{i=1}^{p}\left(Y_i - \sum_{j=1}^{p} Y_j / p\right)^2$ has the chi-square distribution with $p-1$ degrees of freedom then $Y_1,...,Y_p$ are Gaussian random variables with $E(Y_1 - EY_1)^2 = \cdots = E(Y_p - EY_p)^2 = 1$.

Since $E(z) = 0$ and $\text{var}(z) = I$, as an immediate modification of Theorem 1 we have:

**Theorem 2.** The following two statements are equivalent:

(a) $X$ is an ID random vector, the vector $z$ consists of independent components and the random variable $\sum_{i=1}^{p}\left(z_i - \sum_{j=1}^{p} z_j / p\right)^2 = W - \left(\sum_{j=1}^{p} z_j\right)^2 / p$ has the chi-square distribution with $p$-$1$ degrees of freedom.

(b) $X \sim N_p(\mu, \Sigma)$, $p \geq 2$.

**Remark 2.3.** Theorems 1 and 2 treat independent random variables $z_1,...,z_p$. In this context, assuming that $z = (z_1,...,z_p)^T$ is an ID random vector and $z_1,...,z_p$ are from specific ID distributions with finite fourth moments, we have that $z_1,...,z_p$ are independent iff $E(z_i^2 z_j^2) = E(z_i^2) E(z_j^2)$, $i \neq j$ for all $1 \leq i, j \leq p$ (see Pierre, 1971, for details). That is to say, a natural question is when are components of an ID vector independent? In this context, Pierre (1971) and Veeh (1982) discussed necessary and sufficient conditions in a parallel with those available in the normal case. It turns out that if the ID vector has finite fourth moment, then pairwise independence is equivalent to total independence.

**Remark 2.4.** It is clear that the problem considered in Theorems 1 and 2 can be associated with the issue of reconstructing a summands distribution by a distribution of their sum. Even in the simple case of $X \sim N_p(\mu, I)$, it turns out that by virtue of the results of Prokhorov and Ushakov (2002) (see Theorem 1 and its Corollary in the cited paper), the ID restriction on $z$'s distributions cannot be significantly improved. In this context, in a general case, the condition "$z_1,...,z_p$ are independent" seems to be essential.

## 3. TESTS

In this section, we exemplify simple applications of the likelihood projection technique to test for MN, employing available software products. The developed test procedures are experimentally evaluated.

Generally speaking, the univariate likelihood projections can yield simple ways to construct tests for MN, e.g., combining a test for $z^T z \sim \chi_p^2$ with a decision making rule for that the random vector $z$ consists of $p$ independent components. Designs, when test strategies combine statistics with structures based on related paradigms, can significantly simplify the development of the tests for MN. For example, taking into account the schematic rule "Likelihood($z^T z \sim \chi_p^2$, $z_1,...,z_p$ are independent) = Likelihood($z^T z \sim \chi_p^2 / z_1,...,z_p$ are independent)×Likelihood($z_1,...,z_p$ are independent)", one can employ a sum of test statistics that are based on log-likelihood type concepts.

Without loss of generality, we exemplify the proposed approach via testing of bivariate normality. (See Remark 3.1 below for testing of trivariate normality.) To this end, we transform the quadratic form $z^T z = W = (X - \mu)^T A (X - \mu)$ via $J = G(W)$, where $G(x) = \int_0^x \exp(-u/2) du / 2$ is the chi-squared distribution function with 2 degrees of freedom. Then, we can aim to test for $J \sim Unif[0,1]$, assessing that $W \sim \chi_2^2$. In this statement, the smooth Neyman test for uniformity (e.g., Ledwina, 1994), a log-likelihood structured decision making mechanism, uses the statistic

$$T_{1n} = \frac{1}{n} \sum_{j=1}^{k_{1n}} \left( \sum_{i=1}^{n} b_j(J_i) \right)^2,$$

where values $J_1,...,J_n$, independent realizations of $J$, are assumed to be observed; $b_1,...,b_{k_1}$ are normalized Legendre polynomials on [0,1]; and $k_{1n}$ is proposed to be chosen via the data-driven procedure, a modified Schwarz's rule, developed by Ledwina (1994) and Inglot and Ledwina (2006). In order to obtain values of $T_{1n}$, we can employ the R-command (R Development Core Team, 2012): *ddst.uniform.test* that is contained in the R-package 'ddst' (https://cran.r-project.org/web/packages/ddst/ddst.pdf). To test for independence between $z_1$ and $z_2$, we apply the data-driven rank strategy proposed by Kallenberg and Ledwina (1999). The log-likelihood type test statistic is

$$T_{2n} = \frac{1}{n} \sum_{j=1}^{k_{2n}} \left\{ \sum_{i=1}^{n} b_j \left( \frac{R_{1i} - 1/2}{n} \right) b_j \left( \frac{R_{2i} - 1/2}{n} \right) \right\}^2,$$

where we assume that samples $(z_{j1},...,z_{jn})$ related to random variables $z_j$, $j = 1, 2$, are observed; $R_{ji}$ denotes the rank of $z_{ji}$ among $(z_{j1},...,z_{jn})$, $j = 1, 2$; and $k_{2n}$ is chosen in the data-driven manner, a modified Schwarz's rule, shown in Kallenberg and Ledwina (1999). To implement this procedure, we can use the R-command *testforDEP* that is contained in the R-package 'testforDEP' (Miecznikowski et al., 2018). Thus, the test statistic for bivariate normality has the form $T_n = T_{1n} + T_{2n}$.

In practice, the parameters of the null distribution of the vector *X* are unknown. Thus, finally applying a common approach in assessing MN of underlying data distributions based on the residuals (e.g., Baringhaus and Henze, 1988), we obtain the following decision making procedure. Let $_iX = (_iX_1, _iX_2)^T$, $i = 1,...,n$, be independent identically distributed bivariate random vectors that are realizations of $X = (X_1, X_2)^T$, with sample mean $\bar{X}_n = \sum_{i=1}^{n} (_iX)/n$ and sample covariance matrix $S_n = \sum_{i=1}^{n} (_iX - \bar{X})(_iX - \bar{X})^T / n$. Assume $_1X \sim N_2(\mu, \Sigma)$ under the null hypothesis. Then, we can compute $S_n^{-1/2}$ that is (almost surely) the unique symmetric positive-definite square root of the inverse of $S_n$ which is positive-definite with probability one (Eaton and Perlman, 1973). Define the residuals $\tilde{z}_i = (\tilde{z}_{1i}, \tilde{z}_{2i})^T = S_n^{-1/2}(_iX - \bar{X})$ and the statistics $\tilde{J}_i = G(\tilde{z}_i^T \tilde{z}_i)$, $i = 1,...,n$. The null hypothesis is rejected for large values of

$$\tilde{T}_n = \frac{1}{n}\sum_{j=1}^{k_{1n}}\left(\sum_{i=1}^{n}b_j(\tilde{J}_i)\right)^2 + \frac{1}{n}\sum_{j=1}^{k_{2n}}\left\{\sum_{i=1}^{n}b_j\left(\frac{\tilde{R}_{1i}-1/2}{n}\right)b_j\left(\frac{\tilde{R}_{2i}-1/2}{n}\right)\right\}^2,$$

where $\tilde{R}_{ji}$ denotes the rank of $\tilde{z}_{ji}$ among $(\tilde{z}_{j1},...,\tilde{z}_{jn})$, $j=1,2,$, $k_{1n}, k_{2n}$ are chosen in the data-driven manner based on observations $\tilde{J}_i, \tilde{z}_i, i=1,...,n$ (see the $T_{1n}, T_{2n}$-strategies above, respectively). To compute values of the test statistic $\tilde{T}_n$, one can use the R code:

*library(ddst); library(testforDEP); zz<-z1^2+z2^2; J<-pchisq(zz,2); T<-ddst.uniform.test(J, compute.p=FALSE)$statistic+testforDEP(z1,z2,test="TS2",num.MC = 100)@TS*

### 3.1. Null distribution

According to Szkutnik (1987), the null distribution of the residuals based test statistic $\tilde{T}_n$ does not depend on the parameters $(\mu, \Sigma)$ under the null hypothesis (see also, e.g., Baringhaus and Henze, 1988). However Henze (2002) provided concerns regarding this fact. We then present the critical values for the proposed test for different sample sizes using the Monte Carlo technique, and experimentally examine this result for different values of $\rho = cor(_iX_1, _iX_2)$, $i = 1,...,n$.

In order to tabulate the percentiles of the null distribution of the test statistic $\tilde{T}_n$, we drew 55,000 samples of $_1X,...,_nX \sim N_2\left(\begin{bmatrix}0\\0\end{bmatrix}, \begin{bmatrix}1 & -0.5\\-0.5 & 1\end{bmatrix}\right)$ calculating values of $\tilde{T}_n$ at each sample size $n$. The generated values of the test statistic $\tilde{T}_n$ were used to determine the critical values $C_\alpha$ of the null distribution of $\tilde{T}_n$ at the significance levels $\alpha$. The results of this Monte Carlo study are displayed in Table 1.

**Table 1.** *Critical Values of the Proposed Test Statistic*

| | | $\alpha$ | | | | | $\alpha$ | | |
|---|---|---|---|---|---|---|---|---|---|
| **n** | 0.2 | 0.1 | 0.05 | 0.01 | **n** | 0.2 | 0.1 | 0.05 | 0.01 |
| 25 | 0.9759 | 5.5513 | 9.3079 | 18.6332 | 60 | 0.6805 | 1.5788 | 6.7011 | 15.2855 |
| 30 | 0.8577 | 5.2503 | 9.1045 | 18.3795 | 80 | 0.6284 | 1.1314 | 5.9050 | 14.2164 |
| 35 | 0.8017 | 4.8672 | 8.3672 | 16.9611 | 90 | 0.6225 | 1.0960 | 5.9588 | 14.8072 |
| 45 | 0.7256 | 4.4877 | 7.7548 | 16.6368 | 100 | 0.6192 | 1.0682 | 5.8645 | 13.7717 |
| 50 | 0.7069 | 4.2035 | 7.1607 | 16.2087 | 125 | 0.6014 | 0.9961 | 5.5347 | 13.2184 |

In order to verify the results shown in Table 1, for different values of $\rho \in (-1,1)$ and $n$, we calculated the Monte Carlo approximations to

$$Pr\left\{\tilde{T}_n > C_{0.05} / \left(_iX_1, _iX_2\right)^T \sim N_2\left(\begin{pmatrix}0\\0\end{pmatrix}, \begin{pmatrix}1 & \rho\\ \rho & 1\end{pmatrix}\right), i=1,...,n\right\}, j=1,2,$$

where $C_{\alpha=0.05}$'s are shown in Table 1. In this study, we also examined the Shapiro-Wilk test (SW), using the R-procedure "*mvShapiro.Test*". For each value of $\rho$ and $n$, the Type I error rates were derived using 75,000 samples of $\left(_iX_1, _iX_2\right)^T \sim N_2\left(\begin{pmatrix}0\\0\end{pmatrix}, \begin{pmatrix}1 & \rho\\ \rho & 1\end{pmatrix}\right), i=1,...,n$. Table 2 presents the results of this Monte Carlo evaluation.

**Table 2.** *The Monte Carlo Type I error probabilities of the proposed test, $\tilde{T}_n$, and the Shapiro-Wilk test (SW), when $\left(_iX_1, _iX_2\right)^T \sim N_2\left(\begin{pmatrix}0\\0\end{pmatrix}, \begin{pmatrix}1 & \rho\\ \rho & 1\end{pmatrix}\right), i=1,...,n$ and the anticipated significance level is $\alpha = 0.05$.*

| | $n=35$ | | $n=50$ | | $n=100$ | |
|---|---|---|---|---|---|---|
| $\rho$ | $\tilde{T}_n$ | SW | $\tilde{T}_n$ | SW | $\tilde{T}_n$ | SW |
| -0.9 | 0.0505 | 0.0511 | 0.0499 | 0.0501 | 0.0499 | 0.0492 |
| -0.7 | 0.0501 | 0.0515 | 0.0498 | 0.0513 | 0.0500 | 0.0497 |
| -0.5 | 0.0498 | 0.0507 | 0.0500 | 0.0483 | 0.0500 | 0.0488 |
| -0.3 | 0.0501 | 0.0501 | 0.0499 | 0.0489 | 0.0501 | 0.0505 |
| -0.1 | 0.0506 | 0.0507 | 0.0495 | 0.0496 | 0.0496 | 0.0491 |
| 0 | 0.0510 | 0.0512 | 0.0501 | 0.0512 | 0.0500 | 0.0498 |
| 0.1 | 0.0495 | 0.0514 | 0.0502 | 0.0505 | 0.0499 | 0.0500 |
| 0.3 | 0.0500 | 0.0494 | 0.0498 | 0.0493 | 0.0500 | 0.0513 |
| 0.5 | 0.0510 | 0.0510 | 0.0499 | 0.0504 | 0.0499 | 0.0489 |
| 0.7 | 0.0495 | 0.0505 | 0.0507 | 0.0503 | 0.0499 | 0.0482 |
| 0.9 | 0.0497 | 0.0506 | 0.0498 | 0.0488 | 0.0500 | 0.0486 |

According to Table 2, the validity of the critical values related to the test statistic $\tilde{T}_n$ is experimentally confirmed

### 3.2. Power

In general, in the considered goodness-of-fit framework, there are no most powerful decision making mechanisms. We examine the proposed approach in several scenarios, where decisions to reject MN can be anticipated to be difficult. Taking into account that "As recommended by many authors …, a reasonable first step in assessing MVN is to test each variable separately for the univariate normality" (Looney, 1995), we consider the designs displayed in Table 3, where $X_1$ and $X_2$ are normally distributed, whereas $X = (X_1, X_2)^T$ is not $N_2$-distributed.

**Table 3.** *Distributions for $X = (X_1, X_2)^T$ used in the power study*

| Alternative Designs | *Models/Descriptions* |
|---|---|
| A$_1$ | $X_1 = \xi_1$, $X_2 = |\xi_2| I(\xi_1 \geq 0) - |\xi_2| I(\xi_1 < 0)$, where $I(.)$ is the indicator function and $\xi_1, \xi_2$ are independent random variables: $\xi_1 \sim N_1(0,1)$, $\xi_2 \sim N_1(0,1)$ (Stoyanov, 2014: p. 88) |

A$_2$   $X$ is from the two dimensional density function

$$f(x_1, x_2) = \varphi(x_1)\varphi(x_2)\{1 + \varepsilon(2\Phi(x_1) - 1)(2\Phi(x_2) - 1)\} \quad \text{with} \quad \Phi(x) = \int_{-\infty}^{x} \varphi(u) du \quad \text{and}$$

$\varepsilon = 0.999$ (Stoyanov, 2014: p. 89).

A$_3$   $$f(x_1, x_2) = \exp\{-\rho^{-2}(x_1^2 - 2\rho x_1 x_2 + x_2^2)/2\} / \{\pi(1-\rho^2)^{1/2}\} I(x_1 x_2 \geq 0), \rho = 0.9$$

(Stoyanov, 2014: p. 89).

A$_4$   $f(x_1, x_2) = \varphi_1(x_1, x_2)/2 + \varphi_2(x_1, x_2)/2$, where $\varphi_1(x_1, x_2)$ and $\varphi_2(x_1, x_2)$ are standard bivariate normal densities with correlation coefficients $\rho_1 = -0.5$ and $\rho_2 = 0.5$, respectively. In this case, $X_1$ and $X_2$ are uncorrelated (Stoyanov, 2014: p. 93).

A$_5$   $$f(x_1, x_2) = \left[\exp\{-2(x_1^2 + x_1 x_2 + x_2^2)/3\} + \exp\{-2(x_1^2 - x_1 x_2 + x_2^2)/3\}\right] / (2\pi 3^{1/2}).$$ In this case, $X_1$ and $X_2$ are uncorrelated, but dependent (Stoyanov, 2014: p. 93).

A$_6$   $f(x_1, x_2) = \exp\{-(1+x_1^2)(1+x_2^2)\}/C$, where $C \simeq 0.993795$. In this case, all the conditional distributions of $X$ are normal (Stoyanov, 2014: p. 97).

A$_7$   $X = \left(\xi^{1/2}\eta_1 + (1-\xi)^{1/2}\eta_2, \xi^{1/2}\eta_3 + (1-\xi)^{1/2}\eta_2\right)^T$, where $\xi \sim Unif[0,1]$ and $\eta_1, \eta_2, \eta_3$ are independent $N_1(0,1)$ distributed random variables (Stoyanov, 2014: p. 97-98).

Table 4 shows the results of the power evaluations of the proposed test $\tilde{T}_n$, the SW test, Henze-Zirklers's MN test (HZ) and the classical Mardia's MN test (M) via the Monte Carlo study based on 55,000 replications of the independent identically distributed bivariate random vectors $_1X, ..., _nX$ for designs A$_1$-A$_7$ at each sample size $n$. To implement the HZ test (Baringhaus and Henze, 1988), we used the R-procedure *mvn(X,mvnTest="hz")* from the package *MVN*. The R-command *mardia(X,plot=FALSE)* was employed to conduct the M test. The significance level of the tests was fixed at 5%.

**Table 4.** *The Monte Carlo power of the tests.*

| | Design $A_1$ | | | | Design $A_2$ | | | |
|---|---|---|---|---|---|---|---|---|
| **Tests/*n*** | **25** | **50** | **100** | **125** | **25** | **50** | **100** | **125** |
| $\tilde{T}_n$ | 0.194 | 0.639 | 0.979 | 0.998 | 0.057 | 0.062 | 0.096 | 0.120 |
| HZ | 0.171 | 0.447 | 0.946 | 0.991 | 0.049 | 0.056 | 0.064 | 0.071 |
| M | 0.065 | 0.134 | 0.209 | 0.236 | 0.021 | 0.046 | 0.069 | 0.076 |
| SW | 0.175 | 0.304 | 0.558 | 0.676 | 0.056 | 0.057 | 0.058 | 0.061 |
| | Design $A_3$ | | | | Design $A_4$ | | | |
| $\tilde{T}_n$ | 0.201 | 0.501 | 0.875 | 0.945 | 0.075 | 0.142 | 0.335 | 0.434 |
| HZ | 0.141 | 0.376 | 0.778 | 0.888 | 0.065 | 0.079 | 0.101 | 0.114 |
| M | 0.054 | 0.306 | 0.684 | 0.796 | 0.054 | 0.119 | 0.201 | 0.236 |
| SW | 0.036 | 0.049 | 0.096 | 0.131 | 0.044 | 0.046 | 0.045 | 0.047 |
| | Design $A_5$ | | | | Design $A_6$ | | | |
| $\tilde{T}_n$ | 0.073 | 0.145 | 0.337 | 0.444 | 0.153 | 0.401 | 0.840 | 0.921 |
| HZ | 0.067 | 0.077 | 0.102 | 0.109 | 0.092 | 0.205 | 0.458 | 0.571 |
| M | 0.053 | 0.117 | 0.201 | 0.237 | 0.037 | 0.182 | 0.452 | 0.548 |
| SW | 0.043 | 0.041 | 0.046 | 0.045 | 0.048 | 0.053 | 0.062 | 0.066 |

| | Design $A_7$ | | | |
|---|---|---|---|---|
| **Tests/*n*** | **25** | **50** | **100** | **125** |
| $\tilde{T}_n$ | 0.0730 | 0.123 | 0.243 | 0.304 |
| HZ | 0.073 | 0.097 | 0.159 | 0.193 |
| M | 0.051 | 0.110 | 0.187 | 0.218 |
| SW | 0.054 | 0.058 | 0.066 | 0.067 |

This study demonstrates that the likelihood projection based test is significantly superior to the considered classical tests in all scenarios $A_1$-$A_7$. Specifically, the proposed test clearly outperforms the classical tests in terms of the power properties when detecting MN based on vectors with uncorrelated $N_1(0,1)$-distributed components. It seems that the SW test is biased under $A_3$ ($n=25, 50$), $A_4$, $A_5$ and $A_6$ ($n=25$) and inconsistent under design $A_5$. The M test is biased under $A_2$ ($n=25, 50$).

Based on the Monte Carlo results, we conclude that the proposed test exhibits high and stable power characteristics in comparison to the well-known classical procedures.

**Remark 3.1.** Assume, for example, we observe trivariate independent identically distributed vectors $_iX = (_iX_1, _iX_2, _iX_2)^T$, $i=1,...,n$ that are realizations of $X = (X_1, X_2, X_3)^T$. In a similar manner to the bivariate case considered above, we may define the residuals $\tilde{z}_i = (\tilde{z}_{1i}, \tilde{z}_{2i}, \tilde{z}_{3i})^T$. By Remark 2.3, in order to test for $X \sim N_3$, we can construct the test statistic

$$\widehat{T}_n = \frac{1}{n}\sum_{j=1}^{k_{1n}}\left(\sum_{i=1}^n b_j(\widehat{J}_i)\right)^2 + \sum_{(s,r)=(1,2),(1,3),(2,3)}\sum_{j=1}^{k_{srn}}\left\{\sum_{i=1}^n b_j\left(\frac{\widehat{R}_{si}-1/2}{n}\right)b_j\left(\frac{\widehat{R}_{ri}-1/2}{n}\right)\right\}^2/n,$$

where $\widehat{J}_i = G(\tilde{z}_i^T \tilde{z}_i)$, $\widehat{R}_{ji}$ is the rank of $\tilde{z}_{ji}$ among $(\tilde{z}_{j1},...,\tilde{z}_{jn})$, $j=1,2,3$, $k_{1n}, k_{srn}$ are chosen in the data-driven manner based on observations $\widehat{J}_i, \tilde{z}_i, i=1,...,n$. For example, for $n=250$, using 55,000 replications of $_1X,...,_nX$, we computed the critical value $C_\alpha = 7.0513$ of the null distribution of $\widehat{T}_n$ at the significance level $\alpha = 0.05$. The corresponding Monte Carlo powers of the $\widehat{T}_n$-based test and the HZ, M, SW tests were obtained as 0.481 and 0.239, 0.227, 0.069, respectively, when $X = \left(\xi^{1/2}\eta_1 + (1-\xi)^{1/2}\eta_2, \xi^{1/2}\eta_3 + (1-\xi)^{1/2}\eta_2, \xi^{1/2}\eta_4 + (1-\xi)^{1/2}\eta_5\right)^T$, where $\xi \sim Unif[0,1]$ and $\eta_j, j=1,...,5$ are independent $N_1(0,1)$-distributed random variables. In this design, $X_1, X_2$ and $X_3$ are $N_1(0,1)$-distributed, $X_3$, conditionally on $X_1, X_2$, has a normal distribution $N_1(0,1)$, however $X$ cannot have a trivariate normal distribution (Stoyanov, 2014: p. 97-98).

## 4. CONCLUDING REMARKS

This paper established new univariate likelihood based projections of the MN distribution. It can be attractive to release the conditions used in the presented theorems as well as extend and methodize the likelihood based concept to characterize different multivariate distributions (e.g., Costa and Hero, 2002).

We developed a new approach for testing of MN. The proposed procedure is simple and can be easily applied in practice, since reliable software products for performing modules of the likelihood projections based tests for MN are available. Through extensive Monte Carlo simulation studies, we showed that, employing the well-known tests based on univariate observations, we developed the strategy to assess MN that is superior to the classical procedures across a variety of settings when non-MN distributed vectors consist of normal variables. In future studies, many types of corresponding univariate-based plots can be constructed to be both easy to make and simple to use for detecting departures from assumed multivariate distributions. It is hoped that the present paper will convince the readers of the usefulness of multivariate distributions' characterizations via relevant likelihood functions.

**Acknowledgments**

The author thanks Professor Lev B. Klebanov, Charles University, Czech Republic, and Professor Abram Kagan, University of Maryland College Park, USA, for many insightful and helpful discussions regarding the presented results.